\documentclass[aps,prl,twocolumn,groupedaddress,showpacs,floatfix,altaffilletter]{revtex4-1}
\usepackage{graphicx}
\usepackage{grffile}
\usepackage{amsmath}
\usepackage{amssymb}
\usepackage{bm}
\usepackage{color}
\usepackage[dvipsnames]{xcolor}
\usepackage{amsmath,amssymb,amsfonts}
\usepackage{epsfig}
\usepackage{times}
\usepackage[colorlinks,bookmarks=false,citecolor=blue,linkcolor=red,urlcolor=blue]{hyperref}

\newcommand{\fref}[1]{Fig.~\ref{#1}}

\begin{document}

\title{Nodal versus nodeless superconductivity in iso-electronic LiFeP and LiFeAs } 

\author{R.~Nourafkan}
\affiliation{D{\'e}partement de Physique and Regroupement qu\'eb\'ecois sur les matériaux de pointe, Universit{\'e} de Sherbrooke, Sherbrooke, Qu{\'e}bec, Canada J1K 2R1}

%
\begin{abstract}
Nodal superconductivity is observed in LiFeP while its counterpart LiFeAs with similar topology and orbital content of the Fermi surfaces is a nodeless superconductor.  We explain this difference by solving, in the two-Fe Brillouin zone, the frequency-dependent Eliashberg equations with spin-fluctuation mediated pairing interaction. Because of Fermi surface topology details, in LiFeAs all the Fe-$t_{2g}$ orbitals favor a common pairing symmetry. By contrast, in LiFeP the $d_{xy}$ orbital favors a pairing symmetry different from $d_{xz/yz}$ and their competition determines the pairing symmetry and the strength of the superconducting instability: $d_{xy}$ orbital strongly overcomes the others and imposes the symmetry of the superconducting order parameter. The leading pairing channel is a $d_{xy}$-type state with nodes on both hole and electron Fermi surfaces. As a consequence, the $d_{xz/yz}$ electrons weakly pair leading to a reduced transition temperature in LiFeP.

\end{abstract}
\pacs{74.20.Pq, 74.70.Xa, 74.20.Rp}

\maketitle
Identification of the characteristics of the electronic structure that cause nodal or nodeless superconducting gap symmetry has been one of the outstanding problems in the field of iron-based superconductors (FeSCs).~\cite{PhysRevLett.113.027002} For the $1111$ family, it has been suggested that the ratio between nearest-neighbor hopping and next-nearest-neighbor hopping within the $d_{xy}$ orbital~\cite{Note1}
%
, $t^{(1)}_{xy,xy}/t^{(2)}_{xy,xy}$, controls the spin excitation spectra and hence is a possible switch between nodeless high-$T_c$ and nodal low-$T_c$ pairings.~\cite{PhysRevLett.113.027002, PhysRevB.79.224511, PhysRevLett.106.187003, PhysRevB.81.184512} In the strong coupling picture,  the hopping ratio determines the ratio between nearest and next-nearest antiferromagnetic exchange couplings $J_1$ and $J_2$.   While $J_2$ promotes nodeless $s^{+-}$- or $d_{xy}$-wave superconductivity,  $J_1$ induces nodal  $s^{+-}$- or $d_{x^2-y^2}$-wave superconductivity.~\cite{PhysRevLett.113.027002} Iron-based superconductors usually show next nearest-neighbour superconductivity with $J_1/J_2 < 1$. Assuming a nodeless $s^{+-}$-wave at large $J_2$,  increasing $J_1$ frustrates that pairing symmetry and eventually induces nodes and reduces $T_c$.
 
The spin-fluctuation mediated pairing interaction is repulsive, requiring a sign changing gap function. Once the inter-pocket interaction dominates, pairing is driven by scattering between hole and electron pockets, with a sign change of the superconducting (SC) gap between pockets. Otherwise,  pairing is due to a direct interaction within hole or within electron pockets with a sign change of the SC gap, leading to nodes on each pocket.~\cite{PhysRevB.84.224505}  In the $1111$ family, a variation in $t^{(1)}_{xy,xy}$ can cause disappearance of the Fermi surface (FS) around the $M$-point in the unfolded Brillouin zone (BZ) and band reconstruction among Fe-$t_{2g}$ bands with a large variation of the $d_{xy}$ portion.~\cite{PhysRevLett.113.027002,PhysRevB.79.224511} 
The disappearance of this hole pocket, for instance from LiFeAsO to LiFePO, makes inter-pocket scattering weaker and hence gives a relatively more important role to the scattering between electron pockets. This in turn induces nodes on these pockets if the repulsive intra-pocket pairing interaction is strong enough.  

Structural factors, such as Fe-Fe distance and pnictogen (PN) height, or PN-Fe-PN bond angle, have been used as proxy for the hopping ratio.~\cite{Note2}
%
 It is shown that the SC transition temperatures, $T_c$, is correlated with the PN-Fe-PN bond angle and  is maximum for  bond angles closer to the prefect tetrahedron value of $109.47\,^{\circ}$.~\cite{0953-2048-23-5-054013}

The above explanation for the electronic reconstruction due to iso-electronic P doping on the As site and its consequence on superconductivity can be questioned for the $111$ family since  all $111$ compounds have similar topology and orbital content of the FSs but the low-energy quasi-particle excitations in the SC state varies: LiFeP ($T_c\simeq 5$~K)~\cite{PhysRevB.82.014514} has a nodal gap structure~\cite{PhysRevLett.108.047003} in contrast to the fully gapped SC state in LiFeAs ($T_c\simeq 18$~K)~\cite{PhysRevB.78.060505}.
Furthermore,  the Fe-PN-Fe bond angle is $\simeq 108.59\,^{\circ}$ and $\simeq 102.79\,^{\circ}$ in LiFeP and LiFeAs, respectively.~\cite{Note3}
%
The bond angle in LiFeP is close to the bond angle that maximizes $T_c$ in the $1111$ family,  yet LiFeP has a smaller $T_c$ than LiFeAs. It thus remains unclear what parameter controls the nodal/nodeless competition in the $111$ family. This competition is most likely between nodeless $s^{+-}$ and nodal $d_{xy}$ that can be caused by similar spin fluctuations. 

Here we focus on LiFeP  and  LiFeAs of the $111$-family.  
We employ spin-fluctuation mediated pairing by considering both the Fe-$3d$ and P-$3p$ or As-$4p$ orbitals in the two-Fe unit cell. We solve the linearized Eliashberg equations to investigate SC pairing and gap symmetry. We find that the important factor is whether different orbitals collaborate or compete in imposing a SC gap symmetry.
As we will see, spin fluctuations in LiFeP come from better nested FSs for the $d_{xy}$ orbital relative to the $d_{xz/yz}$ orbitals.  Hence, the $d_{xy}$ orbital imposes its preferred pairing symmetry. Consequently, LiFeP exhibits strong $d_{xy}$ and very weak $d_{xz/yz}$ Cooper pairing.

\begin{figure}
\begin{center}
\begin{tabular}{cc}
\includegraphics[width=0.36\linewidth]{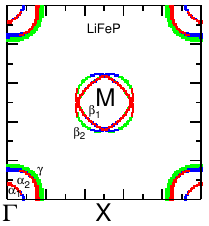} &
\includegraphics[width=0.36\linewidth]{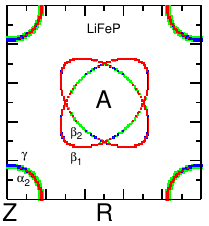}
\\
\includegraphics[width=0.36\linewidth]{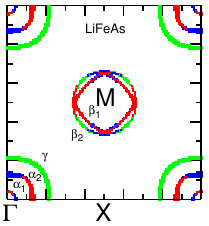} &
\includegraphics[width=0.36\linewidth]{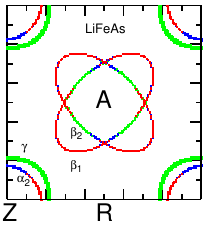}  
\end{tabular}
\end{center}
\caption{(Color online) Partial spectral weight, $A_{ll}({\bf k},0)$, of Fe $t_{2g}$- orbitals on the FS in the $k_x$-$k_y$ plane with $k_z=0$ (left), and  $k_z=\pi/c$ (right) obtained from the LDA calculation for LiFeP
(top panels) and LiFeAs (bottom panels)
. Here the $d_{xy}$, $d_{xz}$, and $d_{yz}$ orbitals are illustrated by green, blue and red colors, respectively. Note that along the diagonal the $d_{xz}$, $d_{yz}$ orbitals have equal weight but in plotting the $d_{xz}$ is masked by  $d_{yz}$. }
\label{fig1}
\end{figure}

\paragraph{Electronic structure}
It is important to discuss first the electronic structure before moving to the charge and spin fluctuations that are at the origin of pairing.  Theoretical studies of FeSCs have shown that the correlation strength depends on the Fe-PN-Fe bond angle and is strongly enhanced when this angle is decreased.~\cite{Nat.Mater.10.1038, PhysRevLett.112.217202} Consistent with this result, transport measurements~\cite{PhysRevB.85.060503} and de Haas-van Alphen study~\cite{PhysRevLett.108.047003} have revealed a weak correlation strength in LiFeP. Hence an LDA calculation should suffice to obtain its electronic structure.~\cite{Note4}
%
The LDA electronic structure calculation shows that the spectral weight at the Fermi energy arises from Fe $t_{2g}$ orbitals $d_{xy}$ and $d_{xz,yz}$. Similar to LiFeAs~\cite{1508.01789}, the FS of LiFeP consists of three hole-like and two electron-like sheets around the center and corners of the BZ respectively. \fref{fig1} illustrates the partial spectral weight, $ A_{ll}({\bf k},0)$, of Fe $t_{2g}$- orbitals on the FSs.  In both compounds, the two inner hole pockets, $\alpha_1$ and $\alpha_2$, are predominantly from $d_{xz}$ and $d_{yz}$ orbitals.  The smallest hole pocket crosses the Fermi level in close vicinity to the $\Gamma$ point. It hybridizes with the $d_{z^2}$ orbital near the $Z$ point and is closed. Both compounds have the large hole-like FS originating purely from in-plane $d_{xy}$ orbitals. The electron pockets, $\beta_{1,2}$, are from an admixture of  $d_{xy}$, $d_{xz}$ and $d_{yz}$ orbitals. They intersect at small $k_z$ and their order flips, i.e., the inner pocket at $k_z=0$ is the outer pocket at $k_z=\pi/c$.

In comparison with LiFeAs, the $\alpha_1$ pocket is bigger and extends further away from the $\Gamma$ point in LiFeP. The two inner hole bands are deeper than the corresponding bands in LiFeAs, while the outer hole band is slightly shallower. This brings middle and outer hole pockets of LiFeP in close vicinity of each other for all $k_z$.  Moreover, bands are also wider in LiFeP consistent with weaker correlation effects. Wider bands lead to different Fermi velocity and  more itinerant electrons and holes in LiFeP. 
\begin{figure}
\includegraphics[width=0.6\columnwidth]{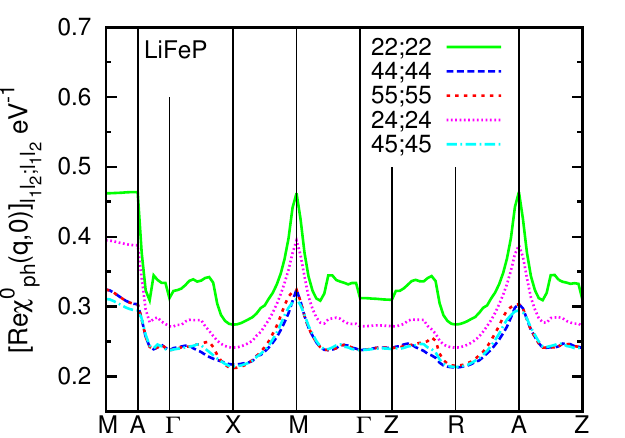}\\
\includegraphics[width=0.6\columnwidth]{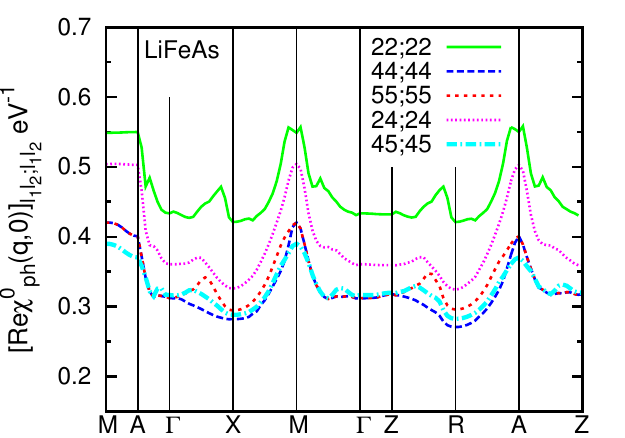}
\caption{(Color online) Comparison between several components of the p-h bare susceptibility
of LiFeP (top panel) and LiFeAs (bottom panel) at $k_BT=0.01$~eV. 
 }\label{fig2}
\end{figure} 

\paragraph{Spin/charge and pairing susceptibilities}
Single-particle excitations in the vicinity of the FSs influence the particle-hole (p-h)  and particle-particle (p-p) susceptibilities involved in pairing.
 Let us begin with bare susceptibilities in the p-h channel  defined as
%
$[{\bm \chi}^{0}_{ph}(Q)]_{l_1l_2;l_3l_4} 
=- (k_BT/N)\sum_K {\bf G}_{K+Q,l_1l_3}{\bf G}_{K,l_4l_2}$,
%
 where ${\bf G}$ denotes the propagator~\cite{Note5}
%
 ,  $l_1,\ldots,l_4$ are combined  ion (or sublattice) and orbital indices and we have defined $K \equiv ({\bf k},\omega_m)$ as momentum-frequency four-vectors and $N$ as the number of points in the BZ. In \fref{fig2} the dominant components of ${\bm \chi}^{0}_{ph}({\bf q},\nu_n=0)$ for LiFeP  are compared with their counterpart in LiFeAs along a high-symmetry path. In what follows, we focus on the Fe-$1$ and Fe-$2$ (on $A$ and $B$ sublattices respectively) $t_{2g}$ orbitals: $d_{xy}$ will be referred as $2$~($7$) and $d_{xz}$ and $d_{yz}$ orbitals as $4$~($9$) and $5$~($10$). As can be seen from \fref{fig2}, all components of the LiFeP susceptibility are smaller than the corresponding components of the LiFeAs susceptibility, which implies that with similar interaction strength, LiFeAs is more unstable toward a magnetic instability. In both compounds, 
 the dominant component of ${\bm \chi}^{0}_{ph}$ is the $d_{xy}$ intra-orbital component, $[{\bm \chi}^{0}_{ph}]_{22;22}$ ($=[{\bm \chi}^{0}_{ph}]_{77;77}=[{\bm \chi}^{0}_{ph}]_{27;27}$), with commensurate (incommensurate) peaks around the $M$ and $A$ points for LiFeP (LiFeAs). These peaks are coming from nesting between the hole and electron pockets. 

Despite their resemblance, the nesting conditions for different orbitals change between the two compounds. In order to quantify this, we introduce the ratio $r\equiv[{\chi}^{0}_{ph}(M-\delta)-{\chi}^{0}_{ph}(\Gamma)]/{\chi}^{0}_{ph}(\Gamma)$ as a measure of nesting between portions of the hole and electron pockets with the same orbital content. $M-\delta$ denotes the momentum of the commensurate or incommensurate peak at the center of the BZ ($\delta$ is zero for LiFeP while it is finite for LiFeAs). This ratio for different orbitals in LiFeP are $r_{d_{xy}} \simeq 0.48$ and $r_{d_{xz/yz}} \simeq 0.35$ while they are  $r_{d_{xy}} \simeq 0.29$ and $r_{d_{xz/yz}} \simeq 0.37$ for LiFeAs. Thus, nesting is better for $d_{xy}$ in LiFeP, while it is better for $d_{xz/yz}$ in LiFeAs.

\begin{figure}
\includegraphics[width=0.6\columnwidth]{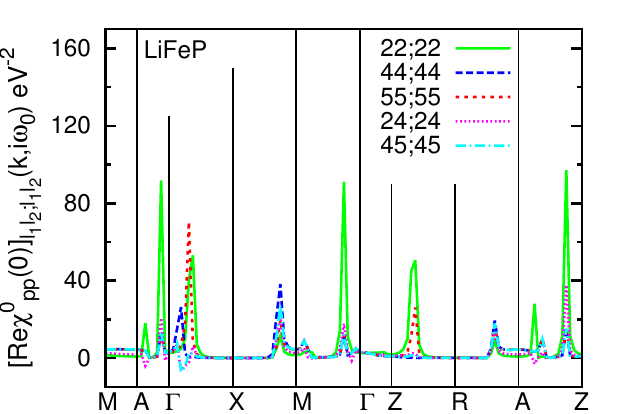}\\
\includegraphics[width=0.6\columnwidth]{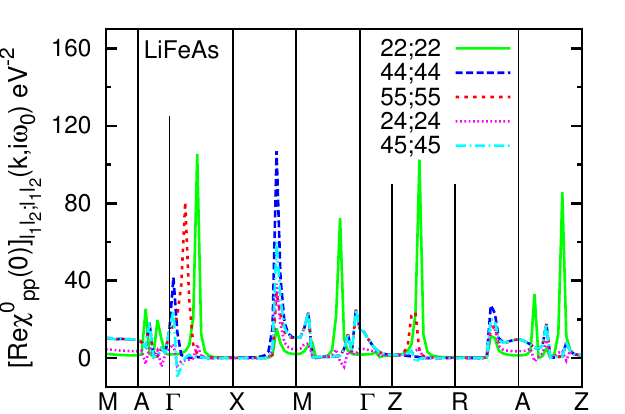}
\caption{(Color online) Real part of several intra-sublattice  components of the generalized p-p bare susceptibility, $(k_BT/N){\bm \chi}^{0}_{pp}(0)$, at the lowest fermionic Matsubara frequency. 
 }\label{fig4}
\end{figure}

The generalized bare susceptibility in the p-p channel is given by
$\left[{\bm \chi}^{0}_{pp}(0)\right]_{K,l_1 l_2;K',l_3 l_4}=
(N/2k_BT){\bf G}_{K,l_1l_3}{\bf G}_{-K,l_2l_4}\delta_{K,K'}$.~\cite{Bickers2004}
\fref{fig4} shows the real part of several components of $(k_BT/N){\bm \chi}^{0}_{pp}(0)$ at the lowest fermionic frequency. The intra-orbital components are purely real and show relatively sharp peaks at the position of FSs. The peak heights are directly proportional to the corresponding orbital weight on the FSs and inversely proportional to the Fermi velocity. They get narrower by reducing temperature.~\cite{Note6}
%
 A close comparison between the two compounds shows that the double peak structure around $M$ for $d_{xz/yz}$ components is strongly reduced in LiFeP. This leads to a suppression of the gap functions for these components. 
The $d_{xy}$ component differs less between the two compounds. Apart from a variation in peak height in the $\Gamma$-$A$ direction,  in LiFeAs the peak in the $\Gamma$-$X$ direction is larger than the peak in $\Gamma$-$M$ direction, which leads to a larger SC gap in this direction on the $\gamma$ pocket.~\cite{1508.01789} The relative magnitude of the two peaks is opposite in LiFeP and one would expect the SC gap to be maximum in the direction toward the $M$-point on the $\gamma$ pocket. This is consistent with a $d_{xy}$-wave or a with a $s^{+-}$-wave gap symmetry with an angle-dependent SC gap which maximizes in this direction. 
It is worth noting that this analysis is possible due to the purely $d_{xy}$ orbital content of this pocket, otherwise orbital content variation and  pairing interaction angular dependence also play roles.

\begin{figure}
\includegraphics[width=0.6\linewidth]{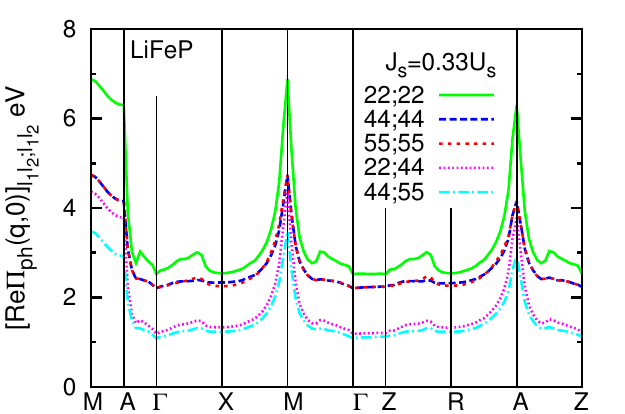}\\
\includegraphics[width=0.6\linewidth]{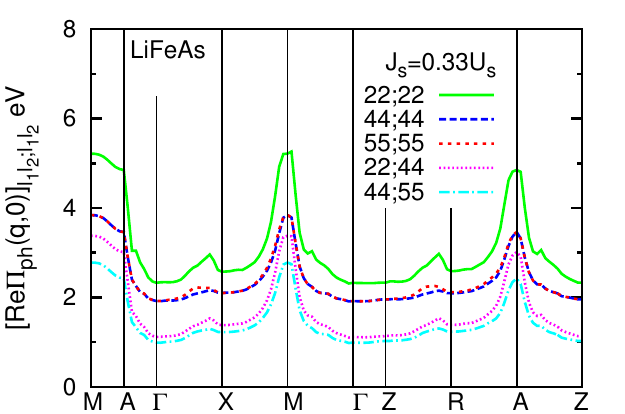}
\caption{(Color online) Comparison between several components of the ladder vertex, ${\bm\Pi}_{ph}$,  
of LiFeP (top panel) and LiFeAs (bottom panel) in the particle-hole channel for two sets of screened interaction parameters that yield the same magnetic Stoner factor: $J_s=0.33U_s, U_s=1.18$~eV for LiFeP and $J_s=0.33U_s, U_s=0.96$~eV for LiFeAs. The inter-orbital interaction and pair hopping are determined assuming spin-rotational symmetry.
}\label{fig3}
\end{figure}

\paragraph{SC instability and pairing interaction}
A SC instability 
occurs when the pairing susceptibility 
diverges as one lowers temperature. The condition for an instability  yields  the linearized Eliashberg equation, defined as (for zero center of mass momentum of Cooper pairs)~\cite{Bickers2004, PhysRevB.57.5376}
\begin{align}
-&(\frac{k_BT}{N})^2\sum_{K'K'',l_3 \ldots l_6}
\left[{\bm \Gamma}^{irr, s}(0)\right]_{K,l_1 l_2;K',l_3 l_4}
\times \nonumber\\
&\left[{\bm \chi}^{0}_{pp}(0)\right]_{K',l_3 l_4;K'',l_5 l_6}
{\bm\Delta}_{K'',l_5l_6} =\lambda(T){\bm \Delta}_{K,l_1l_2},
\label{eq:EL1}
\end{align}
where  ${\bm \Gamma}^{irr, s}$ is the effective pairing interaction in singlet channel and ${\bm \Delta}_{l_1,l_2}({\bf k},i\omega_m)$ is the gap function. In the random phase approximation (RPA),  ${\bm \Gamma}^{irr,s}$, is given by
$[{\bm \Gamma}^{irr,s}(0)]_{Kl_1l_2;K'l_3l_4}=[{\bm \Lambda}^{irr,s}(0)]_{Kl_1l_2;K'l_3l_4}
+[{\bm\Pi}_{ph}(K'-K)]_{l_2l_4;l_3l_1}+[{\bm\Pi}_{ph}(K'+K)]_{l_1l_4;l_3l_2}$,
where the vertex ${\bm \Lambda}^{irr,s}(0)$ is irreducible in all channels.~\cite{Bickers2004, PhysRevB.57.5376, 1508.01789}
The  ladder vertex defined as ${\bm\Pi}_{ph}\equiv-(1/2){\bm\Gamma}^{irr, d}{\bm\chi}_{ph}^d{\bm\Gamma}^{irr, d}+(3/2){\bm\Gamma}^{irr, m}{\bm\chi}_{ph}^m{\bm\Gamma}^{irr, m}$ accounts for the density/magnetic fluctuations contribution in the pairing interaction. Here, 
${\bm \chi}_{ph}^{d(m)}(Q) = {\bm \chi}^{0}_{ph}(Q)/[1
+(-)
{\bm \Gamma}^{irr,d(m)}(Q){\bm \chi}^{0}_{ph}(Q)]$ 
and ${\bm \Gamma}^{irr,d(m)}={\bm \Gamma}^{irr,\uparrow \downarrow}+(-){\bm \Gamma}^{irr,\uparrow \uparrow}$ denote respectively the dressed susceptibility and the irreducible vertex function in the magnetic (density) channel.
In RPA, ${\bm \Gamma}^{irr,\sigma \sigma'}$
is replaced by a static effective vertex which is parametrized by the screened intra-orbital Hubbard interaction, $U_s$, and the Hund's coupling $J_s$ (see SM).~\cite{1508.01789, doi:10.7566/JPSJ.83.043704, PhysRevB.81.054518,PhysRevB.87.045113} 

\fref{fig3} compares the ladder vertex, ${\bm\Pi}_{ph}$, of LiFeP and LiFeAs for material dependent sets of screened interaction parameters that yield the same magnetic Stoner factor ($\alpha^m_S\simeq 0.96$): $J_s=0.33U_s, U_s=1.18$~eV for LiFeP and $J_s=0.33U_s, U_s=0.96$~eV for LiFeAs.~\cite{Note7}
%
Here we only present the intra-sublattice components, which are repulsive and are the dominant terms. 
The most dominant components, i.e. the intra-orbital ones, pair electrons between portions of the FS with the same orbital content.  

 Since the gap equations for different orbitals are coupled by the inter-orbital components of both the pairing interaction and of the bare susceptibility in p-p channel, the competition between the contribution of different orbitals determines the symmetry of the leading gap function. To gain further insight into this competition, we solved the Eliashberg equations for only the $d_{xy}$ orbital and for only the coupled $(d_{xz},d_{yz})$ orbitals. In both compounds, the leading and sub-leading channels for $(d_{xz},d_{yz})$ orbitals are conventional $s^{+-}$ and  $d_{x^2-y^2}$, respectively. This is mainly a consequence of the orbital contents of the FSs illustrated in \fref{fig1}.
However, the leading and sub-leading channels for the $d_{xy}$ orbital are, respectively, $d_{xy}$ and $s^{+-}$ pairing states in LiFeP and $s^{+-}$ and $d_{xy}$ pairing states in LiFeAs. Furthermore, the difference between the corresponding eigenvalues in the two compounds is also larger for the $d_{xy}$ orbital. Note that both $s^{+-}$ and $d_{xy}$ pairing states change sign between portions of the hole and electron pocket which are quasi-nested. Thus, the details of the electronic structure determine which channel is the leading channel. Comparison with a calculation at  $k_BT=0.02$~eV shows that the order of the leading and sub-leading gap symmetries does not change with reducing temperature.

The reasons for this switch of leading and sub-leading channels between the two compounds are related to both nesting and strength of pairing interaction, namely: (i) in LiFeAs the portions of the $\gamma$ pocket around $\theta=0, \pi/2$ are better nested with the portion of the electron pockets that is around the intersection of the two electron pockets, again around $\theta=0, \pi/2$, where $\theta$ is measured at the $\Gamma$ and $M$ points with respect to the $k_x$ axis. This is due to the flatness of the FS on these segments (see SM, Fig.~(1)). This leads to an enhancement of the gap function in these regions, favoring the $s^{+-}$ channel over the $d_{xy}$ channel. In LiFeP, the outer electron FS at $k_z=0$ has a butterfly shape (see SM, Fig.~(1)). The better nested portions are those around $\theta=\pi/4$, leading to a $d_{xy}$ pairing symmetry.  (ii) the already mentioned change in relative strength of the ${\bm \chi}^{0}_{pp}$ peaks at $\Gamma$-$M$ and $\Gamma$-$X$ directions between the two compounds. Therefore, the leading pairing channels for different orbitals are cooperative in LiFeAs while they are competitive in LiFeP. In the full gap equations for LiFeP, the $d_{xy}$ orbital strongly overcomes other orbitals and imposes its symmetry, as we now proceed to show.  In agreement with previous studies,~\cite{PhysRevB.88.174516, PhysRevB.85.014511} gap symmetry of the leading channel of LiFeAs is $s^{+-}$.

\paragraph{SC pairing symmetry of LiFeP}
We obtain the gap function by solving the fully coupled Eliashberg equations. The gap functions are complex and do not change much between $k_z=0$ and $k_z=\pi/c$, hence we present only $k_z=0$ results. The intra-orbital components on the two Fe are equal, while  the intra-sublattice components between one even-parity ($d_{xy}$) and one odd-parity ($d_{xz}$, $d_{yz}$) orbital, change sign between two Fe-ions. Hence the superconducting state does not break parity and glide-reflection symmetry.~\cite{1508.01789} As a consequence of the competition between different Fe-$t_{2g}$ orbitals, the leading pairing channel is a channel with dominant $d_{xy}$ intra-orbital pairing and $B_{2g}$ symmetry.  The inter-sublattice intra-orbital $d_{xy}$ component is the sub-dominant pairing component.  The $d_{xz}$ and $d_{yx}$ intra-orbital components are relatively small and non-degenerate.~\cite{1601.05813} Due to larger $t^{(1)}_{xy,xy}$, LiFeP shows next nearest-neighbor superconductivity  with competing nearest-neighbor Fe-$d_{xy}$ component (see SM for gap function of LiFeP in the orbital basis). By expanding the intra-orbital $d_{xy}$ gap function component in terms of harmonics we find that $\Delta_1 \sin k_x \sin k_y + \Delta_3 (\sin 2k_x \sin k_y+\sin k_x \sin 2k_y)$ is a good approximation.  This channel is the second sub-leading channel in LiFeAs. In short, with the same Stoner factor the $d_{xy}$ Cooper pairs are stronger in LiFeP than in LiFeAs but, due to the weak pairing strength in the $d_{xz/yz}$ orbitals, the overall coherent state is weaker and the transition temperature is lower in LiFeP.

\begin{figure}
\begin{center}
\begin{tabular}{cc}
\includegraphics[width=0.4\columnwidth]{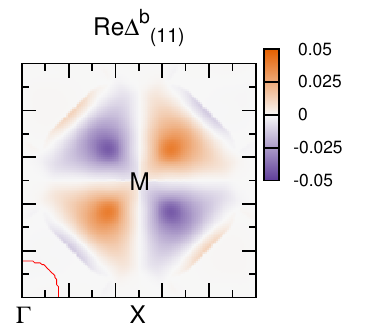} & 
\includegraphics[width=0.4\columnwidth]{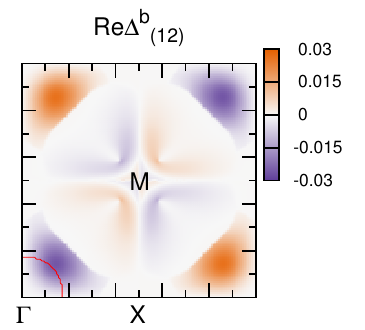} \\
\includegraphics[width=0.4\columnwidth]{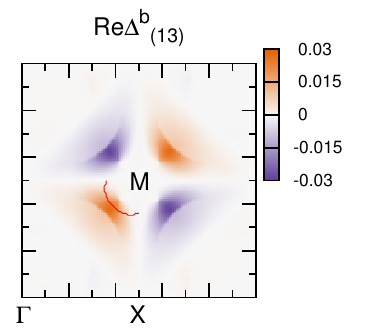} &
\includegraphics[width=0.4\columnwidth]{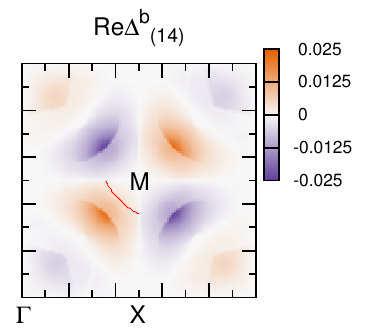} 
\end{tabular}
\caption{(Color online) The real part of the in-plane SC gap function of LiFeP at the lowest Matsubara frequency with largest eigenvalue in the band representation for the $11$th and $12$th (hole) bands (top panels) and the $13$th and
$14$th (electron) bands (bottom panels) for $J_s/U_s=0.33$. The gap function for the inner (hole) band is very small comparatively (not shown).}\label{fig6}
\end{center}
\end{figure}

\fref{fig6} shows the in-plane gap function of LiFeP in the band representation for the $11$th and $12$th ($\alpha_2$ and $\gamma$ hole) bands and the $13$th and
$14$th ($\beta_{1,2}$ electron) bands.~\cite{Note8}
The gap function has nodes on both hole and electron pockets. It is relatively small on the two inner hole pockets (the gap function for $\alpha_1$ hole-band is not shown). The gap function is zero at $\theta = 0, \pi/2$ and increases when
approaching $\theta = \pi/4$ (the direction toward $M$-point) on the
$\gamma$ pocket, while the gap function is maximum at $\theta = \pi/4$ (direction toward $\Gamma$-point) on the $\beta$ pockets and goes to zero when
approaching $\theta = 0, \pi/2$ where the two pockets cross. This is what is expected for $d_{xy}$ pairing symmetry.

\paragraph{Conclusion}
We solved the full linearized Eliashberg gap equation for decoupled $d_{xy}$ and $(d_{xz},d_{yz})$ orbitals and compared the solutions with that of the fully coupled equations in order to distinguish what causes the nodal and nodeless SC gap symmetry observed in LiFeP and LiFeAs respectively.~\cite{PhysRevLett.108.047003}  We find that, in spite of the strong resemblance between the electronic structure of the two compounds, Fe-$t_{2g}$ orbitals all cooperate in LiFeAs with a preferred $s^{+-}$ gap symmetry, whereas in LiFeP the $d_{xy}$ orbital favors $d_{xy}$ pairing symmetry and wins the competition over $d_{xz/yz}$ orbitals that would prefer $s^{+-}$ and hence are weakly paired. This leads to the observed reduction of the transition temperature in LiFeP compared with LiFeAs.

\begin{acknowledgments}
R.~N is deeply indebted to A.-M.S.~Tremblay for many insightful discussions and for careful and critical reading of the manuscript. This work was supported by the Natural Sciences and Engineering Research Council of Canada (NSERC), and by the Research Chair on the Theory of Quantum Materials of Universit\'e de Sherbrooke (A.-M.S.Tremblay). Simulations were performed on computers provided by the Canadian Foundation for Innovation (CFI), the Minist\'ere de l'\'Education du Loisir et des Sports (MELS) (Qu\'ebec), Calcul Qu\'ebec, and Compute Canada.

\end{acknowledgments}
 
%

%
\clearpage
\setcounter{equation}{0}
\setcounter{figure}{0}
\setcounter{table}{0}
\setcounter{page}{1}
\makeatletter
\renewcommand{\theequation}{S\arabic{equation}}
\renewcommand{\thefigure}{S\arabic{figure}}
\renewcommand{\bibnumfmt}[1]{[S#1]}
\renewcommand{\citenumfont}[1]{S#1}
%
\begin{center}
\textbf{\large Supplementary Material: Nodal versus nodeless superconductivity in iso-electronic LiFeP and LiFeAs }
\end{center}
Here we first present some details about the irreducible vertex in RPA. We also present some details about  LDA electronic structure and susceptibility calculations.  Then we compare the $d_{xy}$ partial spectral weights of LiFeP and LiFeAs. Finally, we present some dominant components of the LiFeP gap function in the orbital basis.

\paragraph{Irreducible vertex}
The irreducible vertex function in density/magnetic channels are defined as: ${\bm \Gamma}^{irr,d(m)}={\bm \Gamma}^{irr,\uparrow \downarrow}+(-){\bm \Gamma}^{irr,\uparrow \uparrow}$. In RPA, the irreducible vertex function is replaced by the antisymmetrized static Coulomb vertex, ${\bm \Gamma}^{0,\sigma\sigma'}$ which, in terms of the interacting part of Hamiltonian 
$(1/2)\sum_i\sum_{l_1 \ldots l_4}\sum_{\sigma\sigma'}I^{\sigma\sigma'}_{l_1l_2,l_3l_4}c^{\dagger}_{il_1\sigma}c^{\dagger}_{il_2\sigma'}
c_{il_3\sigma'}c_{il_4\sigma}$ is defined by ${\bm \Gamma}^{0,\sigma \sigma}_{l_1l_2;l_3l_4}=I^{\sigma\sigma}_{l_1l_4,l_3l_2}-I^{\sigma\sigma}_{l_1l_4,l_2l_3}$ and  ${\bm \Gamma}^{0,\sigma \bar{\sigma}}_{l_1l_2;l_3l_4}=I^{\sigma\bar{\sigma}}_{l_1l_4,l_3l_2}$, where $\bar{\sigma}\equiv -\sigma$. For a local interaction the following forms for density/magnetic  irreducible vertex functions are obtained
\begin{align}
{\bm \Gamma}^{irr,d(m)}_{l_1l_2;l_3l_4} = \left\{
  \begin{array}{l l l l l}
    U_s(U_s) & \quad l_1=l_2=l_3=l_4\\
    -U_s'+2J_s(U_s') & \quad l_1=l_3\neq l_2=l_4 \\
    2U_s'-J_s(J_s) & \quad l_1=l_2\neq l_3=l_4 \\
    J_s'(J_s')   & \quad l_1=l_4\neq l_2=l_3 \\
    0         & \quad \text{otherwise}
  \end{array} \right.
  \label{IntPam}
\end{align} 
where $U_s$ and $U_s'$ denote the screened static local intra- and inter-orbital density-density interactions while $J_s$ and $J_s'$ are Hund's coupling and pair-hopping interactions. Due to locality of the interaction the four orbital indices belong to same ion. We also assume spin rotational invariance so the equalities $U_s'=U_s-\frac{5}{2}J_s$~\cite{0034-4885-74-12-124508} and $J_s'=J_s$ are satisfied. Finally, the fully irreducible vertex $[{\bm \Lambda}^{irr,s}(Q)]_{Kl_1l_2;K'l_3l_4}$ is replaced by $\frac{1}{2}\left({\bm \Gamma}^{irr,d}+{\bm \Gamma}^{irr,m}\right)$ transformed to the particle-particle channel.

In our study we used two sets of screened interaction parameters yielding the same magnetic Stoner factor. 
The values for $U_s$ and $J_s$ are relatively  standard  in  the  literature that uses the RPA approach for the pairing vertex.  We found that for a given $J_s/U_s$ ratio changing them within a limited range does not change the qualitative aspects of our results for the superconducting state.

\paragraph{Calculation details}
Both LiFeP and LiFeAs crystallize in a tetragonal structure with a space group $P4/nmm$.  In our study the crystal structures are fixed to the experimental structures. We performed DFT calculations in the Perdew-Burke-Ernzerhof (PBE) scheme using WIEN2k.~\cite{WIEN2k} A very dense ${\bf k}$-point mesh is used in the calculation and the convergence on both charge and energy has reached. An atomic-like basis set is constructed from the Kohn-Sham bands contained within a suitable energy window around the Fermi level for the $d$ shell of Fe, and the $p$ shell of As or P.~\cite{PhysRevB.80.085101} Since, all the orbital with appreciate weight in this energy window are taken into account, we have an orthonormal projection. Then the LDA Hamiltonian is projected from the Kohn-Sham basis to the atomic-like basis. 

For $K$ summations in the definition of the bare susceptibility in p-h channel, we have used a $32\times 32 \times 16$ $\bf k$-mesh and $1024$ positive Matsubara frequencies.  A susceptibility calculation using $(3/2)^3$ more ${\bf k}$-point is done along the high-symmetry path shown in the Fig.~2 of the main text and the results are indistinguishable on the scale of the figure.

\begin{figure}
\begin{center}
\begin{tabular}{cc}
\includegraphics[width=0.45\linewidth]{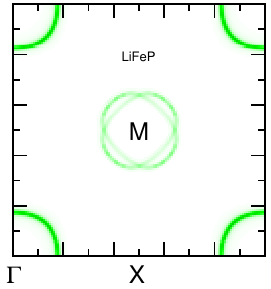} &
\includegraphics[width=0.45\linewidth]{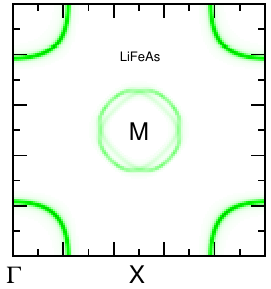} 
\end{tabular}
\end{center}
\caption{(Color online) Partial spectral weight, $A_{xy,xy}({\bf k},0)$, of Fe $d_{xy}$-orbitals on the FS in the $k_x$-$k_y$ plane with $k_z=0$ for LiFeP (left), and  LiFeAs (right). The color scale is the same for both figures. }
\label{fig4-5}
\end{figure}

\paragraph{Partial spectral weight}
\fref{fig4-5} illustrates  the LDA  partial spectral weight, $A_{xy,xy}({\bf k},0)$, of Fe $d_{xy}$-orbitals on the FS in the $k_x$-$k_y$ plane with $k_z=0$ for LiFeP (left), and  LiFeAs (right). While hole- and electron-pocket portions around $\theta=0, \pi/2$ are better nested in LiFeAs, the nesting condition is better around $\theta =\pi/4$ in LiFeP. This can be seen by transferring the hole pocket by a $M-\delta$ vector or by comparing peaks  in the bare (p-p) susceptibility (Fig.~3, main text) in the $\Gamma-X$ and $\Gamma-M$ directions which measure the nesting in the corresponding directions. $M-\delta$ denotes the momentum of the commensurate or incommensurate peak at the center of the BZ ($\delta$ is zero for LiFeP while it is finite for LiFeAs).
 $\theta$ is measured at the $\Gamma$ and $M$ points with respect to the $k_x$ axis.

\begin{figure}
\begin{center}
\begin{tabular}{cc}
\includegraphics[width=0.49\columnwidth]{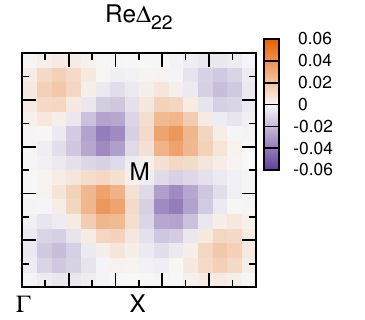} & 
\includegraphics[width=0.49\columnwidth]{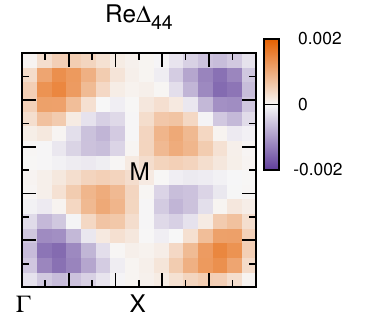} \\
\includegraphics[width=0.49\columnwidth]{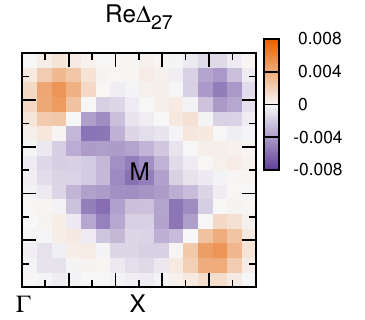} &
\includegraphics[width=0.49\columnwidth]{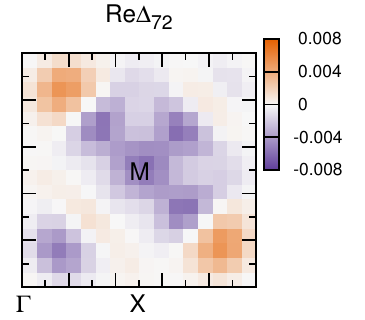} 
\end{tabular}
\caption{(Color online) For $J_s/U_s=0.33$ and $k_BT=0.01$~eV, the real part of the in-plane intra-orbital intra-sublattice (top) and inter-sublattice (bottom) components with the largest eigenvalue in the orbital representation of the LiFeP SC gap function at the lowest Matsubara frequency.}\label{fig5}
\end{center}
\end{figure}

\paragraph{SC gap function of LiFeP in orbital basis}
Finally, \fref{fig5} shows some in-plane intra-orbital intra-sublattice and inter-sublattice components of the gap function in the orbital basis. The $d_{xy}$ intra-orbital intra-sublattice and inter-sublattice components are dominant and sub-dominant components of the gap function. The $d_{xz/yz}$ intra-orbital intra-sublattice components are relatively small leading to a small superconducting gap on the two inner hole pockets (not shown).  The gap function components satisfy the relation $\Delta^{AA(BB)}_{l_1l_2}({\bf k},i\omega_m) = \Delta^{BB(AA)}_{l_1l_2}(-{\bf k},i\omega_m)$ and $\Delta^{AB(BA)}_{l_1l_2}({\bf k},i\omega_m) = \Delta^{BA(AB)}_{l_1l_2}(-{\bf k},i\omega_m)$ which indicate that the superconducting state does not break parity.

\begin{figure}
	\begin{center}
			\includegraphics[width=0.95\linewidth]{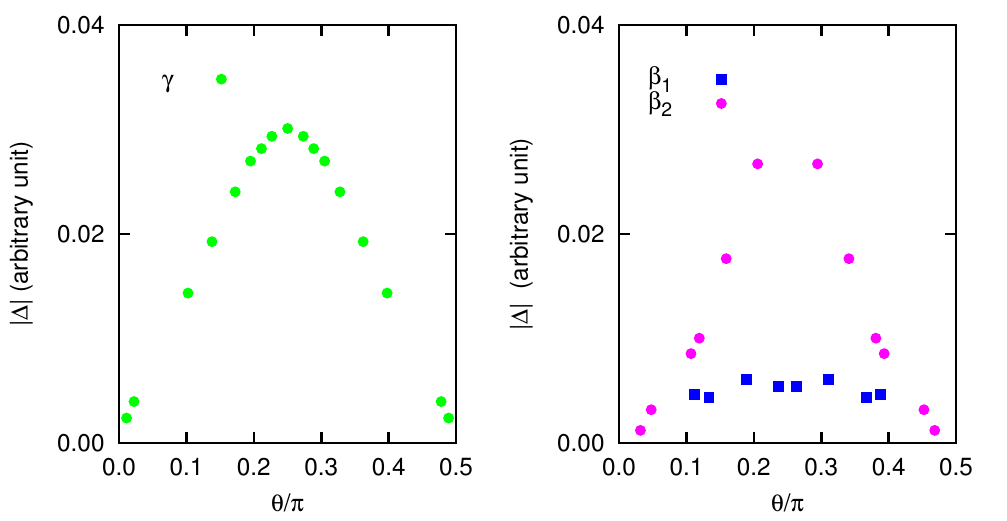} 
	\end{center}
	\caption{(Color online) The SC gap magnitude (in arbitrary unit) as a function of the angle $\theta$  measured at the $\Gamma$ and $M$ points with respect to the $k_x$ axis for $k_z=0$. The SC gap on the $\alpha_{1,2}$ pockets are relatively small.}\label{fig6}
\end{figure}

The linearized Eliashberg gap equation is a dimensionless equation and only gives gap symmetry, not gap magnitude. But one can define a Bogoliubov quasi-particle Hamiltonian and employing the gap function obtained from the gap equation as an estimate of the anomalous self-energy to approximately extract the SC gap on the different FSs. \fref{fig6} demonstrates the angular dependence of the SC gap on $\gamma$ and $\beta_{1,2}$ FSs. The SC gap magnitudes on the $\alpha_{1,2}$ pockets are relatively small (not shown). 
Due to interchange of electron pockets as a function of $k_z$, the gap on the inner pocket becomes larger than that on the outer pocket at a finite $k_z$. Hence, for these pockets, a direct comparison with ARPES data has to take averaging over a range of $k_z$ into account.

\end{document}